\documentclass[12pt,oneside]{article}
\usepackage{epsfig, cite}
\usepackage{color}
\usepackage{ifthen}
\usepackage{graphicx}
\usepackage{amsmath}
\usepackage{txfonts}

\def\({\left(}
\def\){\right)}
\def\[{\left[}
\def\]{\right]}

\def\e{\begin{equation}}
\def\q{\end{equation}}
\def\m{\begin{eqnarray}}
\def\n{\end{eqnarray}}

\begin{document}
\thispagestyle{empty} \setcounter{page}{0}
\renewcommand{\theequation}{\thesection.\arabic{equation}}

\begin{flushright}


\end{flushright}
\vspace{1cm}

\begin{center}
{\huge New Agegraphic Dark Energy in Brans-Dicke Theory}

\vspace{1.4cm}

Xiang-Lai Liu and Xin Zhang

\vspace{.2cm}

{\em Department of Physics, Northeastern
University,\\ Shenyang 110004, China } \\

\vspace{.2cm}



\end{center}

\vspace{0.5cm}

\centerline{ABSTRACT}
\begin{quote}

In this paper, we investigate the new agegraphic dark energy model
in the framework of Brans-Dicke theory which is a natural extension
of the Einstein's general relativity. In this framework the form of
the new agegraphic dark energy density takes as $\rho_{q} =3n^2
\Phi(t)\eta^{-2}$, where $\eta$ is the conformal age of the universe
and $\Phi(t)$ is the Brans-Dicke scalar field representing the
inverse of the time-variable Newton's constant. We derive the
equation of state of the new agegraphic dark energy and the
deceleration parameter of the universe in the Brans-Dicke theory. It
is very interesting to find that in the Brans-Dicke theory the
agegraphic dark energy realizes quintom-like behavior, i.e., its
equation of state crosses the phantom divide $w=-1$ during the
evolution. We also compare the situation of the agegraphic dark
energy model in the Brans-Dicke theory with that in the Einstein's
theory. In addition, we discuss the new agegraphic dark energy model
with interaction in the framework of the Brans-Dicke theory.

\end{quote}
\baselineskip18pt

\noindent

\vspace{5mm}

\newpage

\setcounter{equation}{0}
\section{Introduction}\label{sec:1}

There is no denying that our universe is currently undergoing a
period of accelerated expansion, and consequently the investigation
of dark energy has become one of the hottest topics in modern
cosmology. This cosmic acceleration has been widely proved by many
astronomical observations, especially by the observation of the type
Ia supernovae \cite{ref:a} which provides confirmatory evidence for
this remarkable finding. Combining the analysis of cosmological
observations we realize that the universe is spatially flat and a
mysterious dominant component, dark energy, which is an exotic
matter with large enough negative pressure, leads to this cosmic
acceleration. The preferred candidate of dark energy is the
Einstein's cosmological costant which can fit the observations well,
but is plagued by the ``fine-tuning'' and the ``cosmic coincidence''
problems \cite{Weinberg}. In order to alleviate the
cosmological-constant problems and explain the acceleration
expansion, many dynamical dark energy models have been proposed,
such as quintessence \cite{quint}, phtantom \cite{phantom}, quintom
\cite{quintom}, $k$-essence \cite{k}, hessence \cite{hessence},
tachyon \cite{tachyonic}, Chaplygin gas \cite{Chaplygin}, Yang-Mills
condensate \cite{YMC}, ect.

In fact, the dark energy problem might be in essence an issue of
quantum gravity \cite{Witten}. By far, however, a complete theory of
quantum gravity has not been established, so it seems that we have
to consider the effects of gravity in some effective quantum field
theory in which some fundamental principles of quantum gravity
should be taken into account. It should be stressed that the
holographic principle \cite{Hooft93} is commonly believed as a
fundamental principle of the underlying quantum gravity theory.
Based on the holographic principle, a viable holographic dark energy
model was constructed by Li \cite{Li} by choosing the scale of the
future event horizon of the universe as the infrared cutoff of the
effective quantum field theory. The holographic dark energy model is
very successful in explaining the observational data and has been
studied widely (see, e.g., Refs.~\cite{holoext,intholo,holofit}).
More recently, a new dark energy model, dubbed ``agegraphic dark
energy'' model, has been proposed by Cai \cite{ade}, which is also
related to the holographic principle of quantum gravity. The
agegraphic dark energy takes into account the uncertainty relation
of quantum mechanics together with the gravitational effect in
general relativity.

In the general relativity, one can measure the spacetime without any
limit of accuracy. However, in the quantum mechanics, the well-known
Heisenberg uncertainty relation puts a limit of accuracy in these
measurements. Following the line of quantum fluctuations of
spacetime, K\'{a}rolyh\'{a}zy and his collaborators~\cite{r1} (see
also Ref.~\cite{r2}) made an interesting observation concerning the
distance measurement for Minkowski spacetime through a light-clock
{\it Gedanken experiment}, namely, the distance $t$ in Minkowski
spacetime cannot be known to a better accuracy than
\begin{equation}
\delta t=\lambda t_p^{2/3}t^{1/3}~,\label{eq1}
\end{equation}
where $\lambda$ is a dimensionless constant of order unity.

The K\'{a}rolyh\'{a}zy relation~(\ref{eq1}) together with the
time-energy uncertainty relation enables one to estimate a quantum
energy density of the metric fluctuations of Minkowski
spacetime~\cite{r2,r3}. Following Refs.~\cite{r2,r3}, with respect
to the Eq.~(\ref{eq1}) a length scale $t$ can be known with a
maximum precision $\delta t$ determining thereby a minimal
detectable cell $\delta t^3\sim t_p^2 t$ over a spatial region
$t^3$. Such a cell represents a minimal detectable unit of spacetime
over a given length scale $t$. If the age of the Minkowski spacetime
is $t$, then over a spatial region with linear size $t$ (determining
the maximal observable patch) there exists a minimal cell $\delta
t^3$ the energy of which due to time-energy uncertainty relation
cannot be smaller than~\cite{r2,r3}
\begin{equation}
E_{\delta t^3}\sim t^{-1}~.\label{eq2}
\end{equation}
Therefore, the energy density of metric fluctuations of Minkowski
spacetime is given by~\cite{r2,r3}
\begin{equation}
\rho_q\sim\frac{E_{\delta t^3}}{\delta t^3}\sim \frac{1}{t_p^2
t^2}\sim\frac{M_p^2}{t^2}~,\label{eq3}
\end{equation}
where $M_p$ is the reduced Planck mass. This energy density can be
viewed as the density of dark energy (i.e., the agegraphic dark
energy). Thus, furthermore, the energy density of the agegraphic
dark energy can be written as \cite{ade}
\begin{equation}
\rho_{q} =3n^{2}M^{2}_{p} t^{-2},\label{eq2}
\end{equation}
where the numerical factor $3n^{2}$ is introduced to parameterize
some uncertainties, such as the species of quantum fields in the
universe, the effect of curved spacetime (since the energy density
is derived for Minkowski spacetime) and so on.

In the original version of the agegraphic dark energy model
\cite{ade}, the time scale $t$ in Eq.~(\ref{eq2}) is chosen to be
the age of the universe $T$, however, unfortunately this version
suffers from some internal inconsistencies of the model
\cite{Neupane:2007ra,nade}. To avoid these internal inconsistencies,
a new version of this model was proposed by Wei and Cai \cite{nade}
by replacing the cosmic age $T$ with the cosmic conformal age $\eta$
for the time scale in Eq.~(\ref{eq2}). So, the new agegraphic dark
energy has the energy density \cite{nade}
\begin{equation}
\rho_{q} =3n^{2}M^{2}_{p} \eta^{-2},\label{nade}
\end{equation}
where
\begin{equation}
\eta\equiv\int_0^t\frac{dt}{a}=\int_0^a\frac{da}{a^{2}H}\label{eq3}
\end{equation}
is the conformal age of the universe. The new agegraphic dark energy
model is successful in fitting the observational data and has been
studied extensively \cite{agequint,ageext,INADE}. In this note, we
will study the new agegraphic dark energy model in the framework of
the Brans-Dicke theory.

The Brans-Dicke theory \cite{BD} is a natural alternative and a
simple extension of the Einstein's general relativity theory. Also,
it is the simplest example of a scalar-tensor theory of gravity
\cite{ST}. In the Brans-Dicke theory, the purely metric coupling of
matter with gravity is preserved, thus the universality of free fall
(equivalence principle) and the constancy of all non-gravitational
constants are ensured. The Brans-Dicke theory can pass the
experimental tests from the solar system \cite{solar} and provide an
explanation of the accelerated expansion of the universe
\cite{BDcosmo}. Recently, Wu et al. \cite{Wu1,Wu2} developed the
covariant cosmological perturbation formalism in the case of
Brans-Dicke gravity, and applied this method to the calculation of
cosmic microwave background anisotropy and large scale structures.
Furthermore, they derived observational constraint on the
Brans-Dicke theory in a flat Friedmann-Robertson-Walker (FRW)
universe with the latest Wilkinson Microwave Anisotropy Probe (WMAP)
and Sloan Digital Sky Survey (SDSS) data \cite{Wu2}. In the
Brans-Dicke theory, the gravitational constant is replaced with the
inverse of a time-dependent scalar field, namely,
$\Phi(t)=\frac{1}{8\pi G}$, and this scalar field couples to gravity
with a coupling $\omega$.

Since the Brans-Dicke theory is an alternative to the general
relativity and evokes wide interests in the modern cosmology, it is
worthwhile to discuss dark energy models in this framework.
Recently, the holographic dark energy model has been studied in the
framework of the Brans-Dicke theory \cite{BDholo} (for the case of
the holographic Ricci dark energy, see \cite{BDricci}). In this
work, we consider the agegraphic dark energy model in the
Brans-Dicke theory. We are interested in how the agegraphic dark
energy evolves in the universe in the framework of Brans-Dicke
theory.

This paper is organized as follows: In Sec.~\ref{sec:2}, we briefly
review the Friedmann equation in the Brans-Dicke theory. We study
the original agegraphic dark energy and the new agegraphic dark
energy in Brans-Dicke theory in Secs.~\ref{sec:3} and \ref{sec:4},
respectively. In Sec.~\ref{sec:5}, we further discuss the new
agegraphic dark energy with interaction. In Sec.~\ref{sec:6}, we
give the conclusion.

\setcounter{equation}{0}
\section{Friedmann Equation in Brans-Dicke Theory}\label{sec:2}

First, we will briefly review the Friedmann equation of the
Brans-Dicke cosmology. In the Jordan frame, the action for the
Brans-Dicke theory with matter fields is written as
\begin{equation}\label{BD action}
S = \int d^4x\sqrt{g}\left[ \frac{1}{2}\left( \Phi R - \omega
\frac{\nabla_\mu \Phi \nabla^\mu\Phi}{\Phi} \right) + \mathcal
{L}_M\right]
\end{equation}
where $\Phi$ is the Brans-Dicke scalar field, $\omega$ is the
generic dimensionless parameter of the Brans-Dicke theory, and
${\cal L}_M$ is the Lagrangian of matter fields. In the Jordan
frame, the matter minimally couples to the metric and there is no
interaction between the scalar field $\Phi$ and the matter fields.
The equations of motion for the metric $g_{\mu\nu}$ and the
Brans-Dicke scalar field $\Phi$ are
\begin{equation}
    \begin{split}
       G_{\mu\nu} = R_{\mu\nu} - {1\over 2} g_{\mu\nu}R &= \frac{1}{\Phi} T^M_{\mu\nu}+ T^{BD}_{\mu\nu},\\
       \nabla_\mu \nabla^\mu \Phi &= \frac{1}{2\omega + 3}T^{M \mu}_{\quad
       \mu},
    \end{split}
\end{equation}
where $T^M_{\mu\nu} = (2/\sqrt{g}) \delta(\sqrt{g}\mathcal{L}_M) /
\delta g^{\mu\nu}$ is the energy-momentum tensor for the matter
fields defined as usual, and in cosmology it can be expressed as the
form of perfect fluid
\begin{equation}
T^M_{\mu\nu} = (\rho_M + p_M)U_{\mu}U_{\nu} + p_M g_{\mu\nu},
\end{equation}
where $\rho_M$ and $p_M$ denote the energy density and pressure of
the matter, respectively, and $U_\mu$ is the four velocity vector
normalized as $U_\mu U^\mu = -1$. The energy-momentum tensor of the
Brans-Dicke scalar field $\Phi$ is expressed as
\begin{equation}
T^{BD}_{\mu\nu} = \frac{\omega}{\Phi^2}\left(\nabla_\mu \Phi
\nabla_\nu\Phi - {1\over 2} g_{\mu\nu}\nabla_\alpha \Phi
\nabla^\alpha \Phi \right) + \left(\nabla_\mu \nabla_\nu\Phi -
g_{\mu\nu}\nabla_\alpha \nabla^\alpha \Phi\right)
\end{equation}
Note that in the $\omega \rightarrow \infty$ limit of the
Brans-Dicke theory, the Einstein's general relativity will be
recovered.

Consider now a spatially flat FRW universe containing matter
component and agegraphic dark energy. For simplicity, we also assume
that the Brans-Dicke scalar field is only a time-dependent function,
namely, $\Phi=\Phi(t)$. We can get the equations describing the
background evolution
\begin{equation}
H^2+H\frac{\dot{\Phi}}{\Phi}-\frac{\omega}{6}\frac{\dot{\Phi}^2}{\Phi^2}=\frac{\rho_{m}+\rho_{q}}{
3\Phi},\label{eq9}
\end{equation}
\begin{equation}
2\frac{\ddot{a}}{a}+H^2+\frac{\omega}{2}\frac{\dot{\Phi}^2}{\Phi^2}+2H\frac{\dot{\Phi}}{\Phi}+\frac{\ddot{\Phi}}{\Phi}
=-\frac{p_{q}}{\Phi},\label{eq10}
\end{equation}
where $H={\dot{a}/a}$ is the Hubble parameter, a dot denotes the
derivative with respect to the cosmic time, $\rho_{m}$ is matter
density, $\rho_{q}$ is the density of agegraphic dark energy and
$p_{q}$ is the pressure of agegraphic dark energy. Assuming
$\Phi(t)=\Phi_0a(t)^{\alpha}$ (here we have taken $a_{0}=1$, where
the subscript $0$ denotes the present day), the Friedmann
equation~(\ref{eq9}) becomes
\begin{equation}
H^2(1+\alpha-\frac{\omega\alpha^2}{6})=\frac{\rho_{m}+\rho_{q}}{3\Phi}.\label{eq13}
\end{equation}
It is easy to see from the Friedmann equation~(\ref{eq13}) in
Brans-Dicke theory that in the limit of $\alpha \rightarrow0$ the
standard cosmology will be recovered.

\setcounter{equation}{0}
\section{The Old Version: Age of the Universe as Time
Scale}\label{sec:3}

We shall first consider the old version of the agegraphic dark
energy model. In this version, the time scale is chosen as the age
of the universe,
\begin{equation}
T=\int_{0}^{a}\frac{da}{aH},\label{eq15}
\end{equation}
so in the Brans-Dicke theory the energy density of the agegraphic
dark energy is given by
\begin{equation}
\rho_{q} =3n^2 \Phi(t)T^{-2}.\label{eq14}
\end{equation}

The Friedmann (\ref{eq13}) can be rewritten as
\begin{equation}
 H^2=H^2_0\Omega_{m0}a^{-(3+\alpha)}+\Omega_{q}H^2,\label{eq16}
\end{equation}
where $\Omega_{m0}=\frac{2}{(6+6\alpha-\omega
\alpha^2)}\frac{\rho_{m0}}{\Phi_{0} H^2_{0}}$ and
\begin{equation}
\Omega_{q}=\frac{2}{6+6\alpha-\omega
\alpha^2}\frac{1}{\Phi}\frac{\rho_{q}}{H^2}=\tilde{n}_{0}\frac{1}{H^2
T^2}\label{eq17}
\end{equation}
with $\tilde{n}_{0}=\frac{6n^2}{6+6\alpha-\omega\alpha^2}$. Using
Eqs.~(\ref{eq15}) and (\ref{eq17}), we get
\begin{equation}
\int_{0}^{a}\frac{da}{aH}=\frac{1}{H}\sqrt{\frac{\tilde{n}_{0}}{\Omega_{q}}}.\label{eq18}
\end{equation}
From Eq.~(\ref{eq16}), we obtain
\begin{equation}
\frac{1}{H}=\sqrt{a^{(3+\alpha)}(1-\Omega_{q})}\frac{1}
{H_{0}\sqrt{\tilde{n}_{0}}}.\label{eq19}
\end{equation}
Considering Eqs.~(\ref{eq18}) and (\ref{eq19}), we obtain the
eqution of motion for $\Omega_{q}$ as
\begin{equation}
\Omega_{q}'=\Omega_{q}\left(1-\Omega_{q}\right)\left(3+\alpha-\frac{2}{\sqrt{\tilde{n}_{0}}}\sqrt{\Omega_{q}}\right),\label{eq20}
\end{equation}
where the prime denotes the derivative with respect to $\ln a$. The
energy conservation equation $ \dot{\rho}_q+3H(1+w_q)\rho_q=0$ leads
to
\begin{equation}
w_{q}=-1-\frac{1}{3}\frac{d \ln \rho_{q}}{d \ln a}.\label{eq21}
\end{equation}
Using Eqs.~(\ref{eq20}), (\ref{eq21}) and
$\rho_{q}=\frac{\Omega_{q}}{1-\Omega_{q}}\rho_{m0}a^{-3}$, we get
the equation of state of the old agegraphic dark energy
\begin{equation}
w_{q}=-\frac{1}{3}\left(3+\alpha-\frac{2}{\sqrt{\tilde{n}_{0}}}\sqrt{\Omega_{q}}\right).\label{eq22}
\end{equation}
Now let us consider the current value of the Brans-Dicke scalar
field $\Phi$. We naturally have $\Phi_0=1/8\pi G$ today, and
obviously we can let $\Omega_{m0}=\frac{2}{(6+6\alpha-\omega
\alpha^2)}\frac{\rho_{m0}}{\Phi_{0} H^2_{0}}\equiv\frac{8\pi
G\rho_{m0}}{3H_{0}^{2}}$. Thus, the relations $\omega\alpha=6$ and
$\tilde{n}_{0}=n^2$ are derived. This leads to Eqs.~(\ref{eq20}) and
(\ref{eq22}) rewritten as
\begin{equation}
\Omega_{q}'=\Omega_{q}\left(1-\Omega_{q}\right)\left(3+\alpha-\frac{2}{n}\sqrt{\Omega_{q}}\right),\label{eq23}
\end{equation}
\begin{equation}
w_{q}=-\frac{1}{3}\left(3+\alpha-\frac{2}{n}\sqrt{\Omega_{q}}\right).\label{eq24}
\end{equation}

Also, we can obtain the deceleration parameter
\begin{equation}
q=-\frac{\ddot{a}}{aH^2}=\frac{3w_{q}\Omega_{q}+1+4\alpha+\alpha^{2}}{2+\alpha}.\label{eqq}
\end{equation}

The solar-system experiments give the result for the value of
$\omega$ is $|\omega|>40000$ \cite{solar}. However, when probing the
larger scales, the limit obtained will be weaker than this result.
In Ref.~\cite{BDobs}, the authors found that $\omega$ is smaller
than 40000 on a cosmological scale. Specifically, Wu and Chen
\cite{Wu2} obtained the observational constraint on the Brans-Dicke
model in a flat universe with cosmological constant and cold dark
matter using the latest WMAP and SDSS data. They found that within
2$\sigma$ range, the value of $\omega$ satisfies $\omega<-120.0$ or
$\omega>97.8$ \cite{Wu2}. They also obtained the constraint on the
rate of change of G at present \cite{Wu2},
\begin{equation}
-1.75\times10^{-12}{\rm yr}^{-1}<\frac{\dot{G}}{G}<
1.05\times10^{-12}{\rm yr}^{-1}
\end{equation}
at $2\sigma$ confidence level. So in our case we get
\begin{equation}
\left|\frac{\dot{G}}{G}\right|=\left|\frac{\dot{\Phi}}{\Phi}\right|=\alpha
H<10^{-12}{\rm yr}^{-1},
\end{equation}
and it implies
\begin{equation}
\alpha<\frac{1}{H}\times 10^{-12}{\rm yr}^{-1}.
\end{equation}
Note that we only consider the positive sector of $\omega$. Taking
the current value of the Hubble constant $h\simeq 0.7$ into account,
one can estimate the bounds on $\alpha$,
\begin{equation}
\alpha<0.01.
\end{equation}
Note also that in Ref.~\cite{Xu:2009ss} Xu, Lu and Li have performed
observational constraints on the holographic dark energy model in
Brans-Dicke theory and they found that the 1$\sigma$ bound on
$\alpha$ is $\alpha<0.14$.

\begin{center}
\begin{figure}[htbp]
\centering
\includegraphics[width=0.49\textwidth]{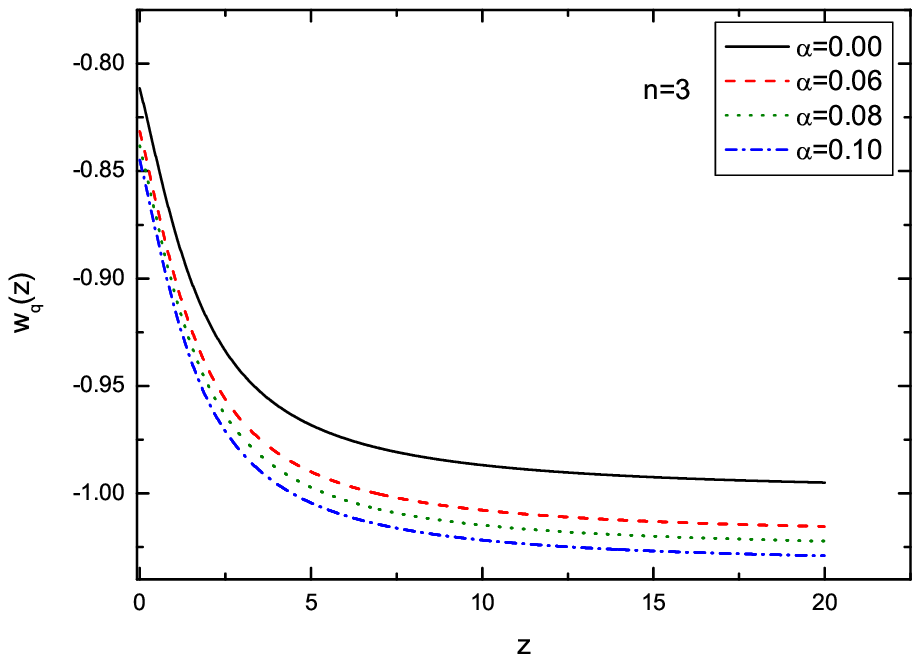}\hfill
\includegraphics[width=0.49\textwidth]{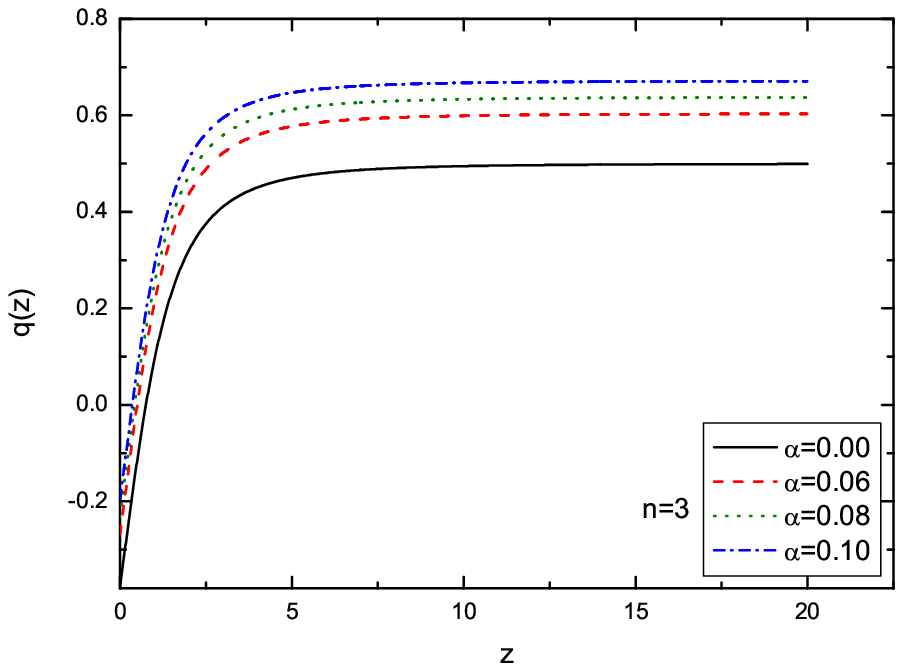}
\caption{\label{fig1} \small The old agegraphic dark energy model in
Brans-Dicke theory: the equation of state of dark energy $w_q(z)$
and the deceleration parameter of the universe $q(z)$. In this
figure, we fix $n=3$ and vary $\alpha$ to compare the usual case
with the Brans-Dicke ones. We take $\Omega_{q0}=0.72$.}
\end{figure}
\end{center}


Figure \ref{fig1} shows the equation of state of dark energy
$w_q(z)$ and the deceleration parameter of the universe $q(z)$ in
the old agegraphic dark energy model within the framework of
Brans-Dicke theory. To compare the usual case ($\alpha=0$) with the
Brans-Dicke ones ($\alpha\neq 0$), we fix $n=3$ and vary $\alpha$ in
this figure. Note that to make a clear comparison we choose some
large values for $\alpha$. Recall that in the usual case the old
agegrahic dark energy behaves like a thawing quintessence
\cite{agequint}. However, in the Brans-Dicke theory, the equation of
state of the old agegraphic dark energy can cross the
cosmological-constant boundary $w=-1$ (``phantom divide''),
realizing the quintom behavior. This can be clearly seen from
Eq.~(\ref{eq24}).

\setcounter{equation}{0}
\section{The New Version: Conformal Age of the Universe as Time
Scale}\label{sec:4}

We now consider the new version of the agegraphic dark energy model
with the dark energy density
\begin{equation}
\rho_{q} =3n^2\Phi(t)\eta^{-2},\label{eq25}
\end{equation}
where $\eta$ is the conformal age of the universe given by
Eq.~(\ref{eq3}).

The Friedmann equation~(\ref{eq13}) can be rewritten as
\begin{equation}
H^2=H^2_0\Omega_{m0}a^{-(3+\alpha)}+\Omega_{q}H^2,\label{eq26}
\end{equation}
where
\begin{equation}
\Omega_{q}=\frac{2}{6+6\alpha-\omega
\alpha^2}\frac{1}{\Phi}\frac{\rho_{q}}{H^2}=\tilde{n}_{0}\frac{1}{H^2
\eta^2}.\label{eq27}
\end{equation}
Combining Eqs.~(\ref{eq3}) and (\ref{eq27}), we obtain
\begin{equation}
\int\frac{da}{a^2H}=\frac{1}{H}\sqrt{\frac{\tilde{n_{0}}}{\Omega_{q}}}.\label{eq28}
\end{equation}
Considering Eqs.~(\ref{eq28}) and (\ref{eq26}), we also obtain the
equation of motion for $\Omega_{q}$,
\begin{equation}
\Omega_{q}'=\Omega_{q}\left(1-\Omega_{q}\right)\left(3+\alpha-\frac{2}{a\sqrt{\tilde{n}_{0}}}\sqrt{\Omega_{q}}\right).\label{eq29}
\end{equation}
Furthermore, linking the conversation equation of dark energy with
$\rho_{q}=\frac{\Omega_{q}}{1-\Omega_{q}}\rho_{m0}a^{-3}$ gives rise
to
\begin{equation}
w_{q}=-\frac{1}{3}\left(3+\alpha-\frac{2}{a\sqrt{\tilde{n}_{0}}}\sqrt{\Omega_{q}}\right).\label{eq30}
\end{equation}

As discussed above, we have the relations $\omega\alpha=6$ and
$\tilde{n}_{0}=n^2$, so Eqs.~(\ref{eq29}) and (\ref{eq30}) are
rewritten as
\begin{equation}
\Omega_{q}'=\Omega_{q}\left(1-\Omega_{q}\right)\left(3+\alpha-\frac{2}{an}\sqrt{\Omega_{q}}\right),\label{eq31}
\end{equation}
\begin{equation}
w_{q}=-\frac{1}{3}\left(3+\alpha-\frac{2}{na}\sqrt{\Omega_{q}}\right).\label{eq32}
\end{equation}
Obviously, the above two equations can also be rewritten as
\begin{equation}
\frac{d\Omega_{q}}{dz}=-\Omega_{q}\left(1-\Omega_{q}\right)\left[(3+\alpha)\frac{1}{1+z}-\frac{2}{n}\sqrt{\Omega_{q}}\right],\label{eq33}
\end{equation}
\begin{equation}
w_{q}=-\frac{1}{3}\left[(3+\alpha)-\frac{2}{n}\sqrt{\Omega_{q}}(1+z)\right].\label{eq34}
\end{equation}
When considering the range of the fractional dark energy density
$0\leq\Omega_{q}\leq1$, one can see from Eq.~(\ref{eq34}) that the
equation of state of the new agegraphic dark energy is in the range
\begin{equation}
-\frac{1}{3}\left(3+\alpha\right)<w_{q}<-\frac{1}{3}\left(3+\alpha-\frac{2}{n}(1+z)\right),\label{eq35}
\end{equation}
so it is obvious that in the Brans-Dicke gravity case the new
agegraphic dark energy can realize the quintom behavior, i.e., the
equation of state of dark energy can cross $-1$ during the
evolution. The deceleration parameter $q$ in this case has the same
form as Eq.~(\ref{eqq}).

\begin{center}
\begin{figure}[htbp]
\centering
\includegraphics[width=0.32\textwidth]{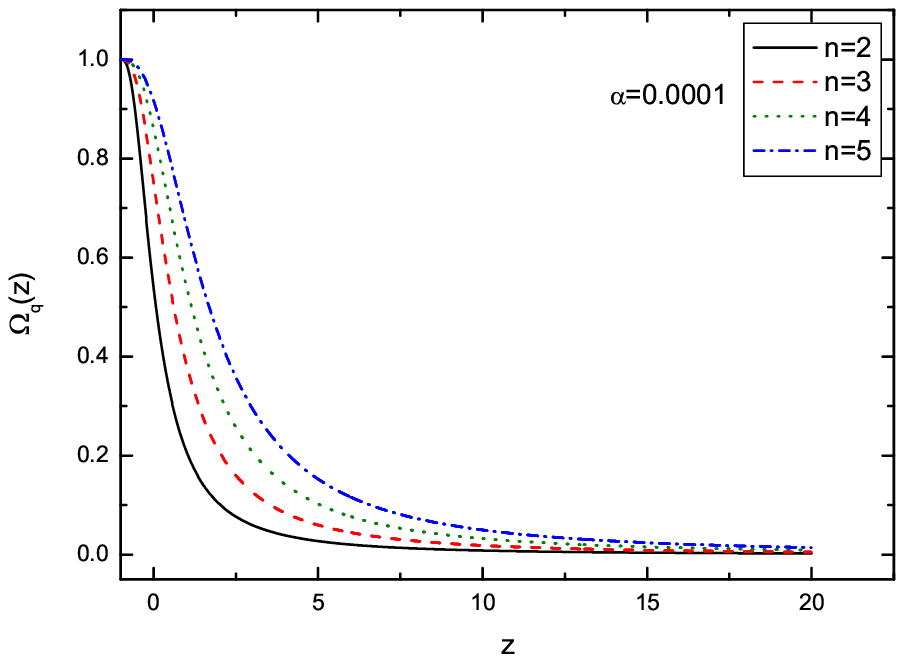}\hfill
\includegraphics[width=0.32\textwidth]{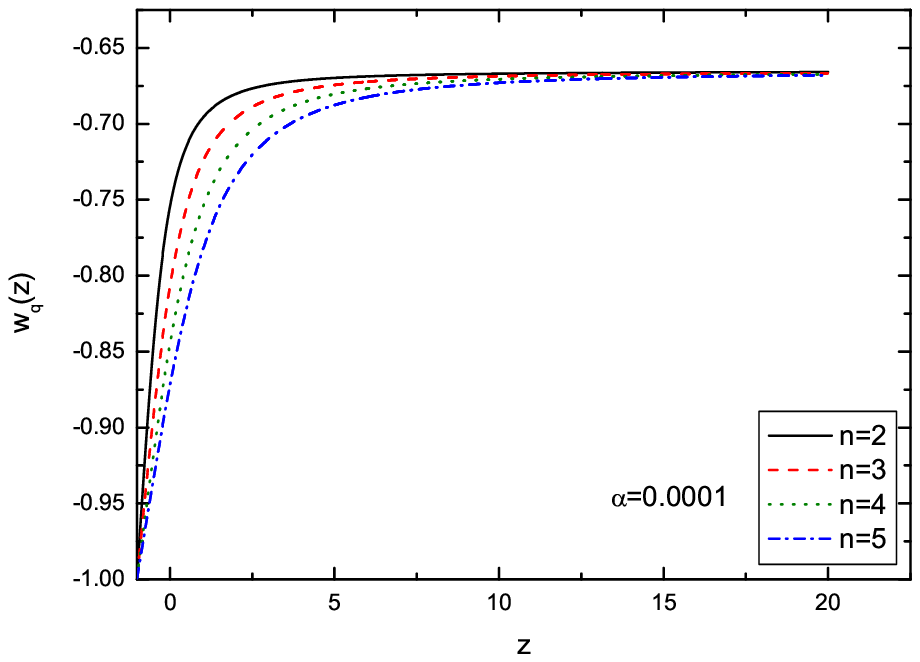}
\includegraphics[width=0.32\textwidth]{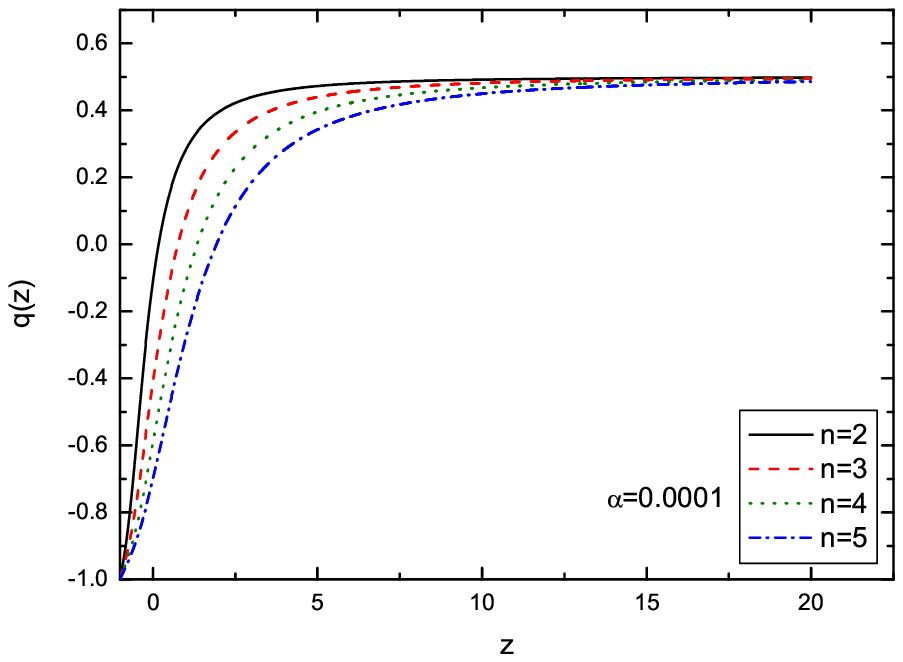}
\includegraphics[width=0.32\textwidth]{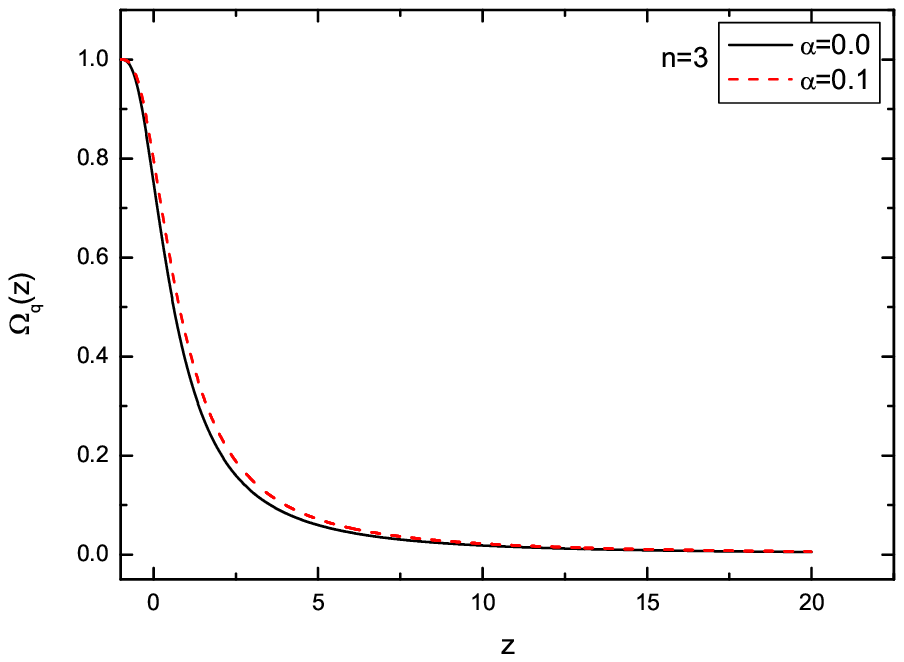}\hfill
\includegraphics[width=0.32\textwidth]{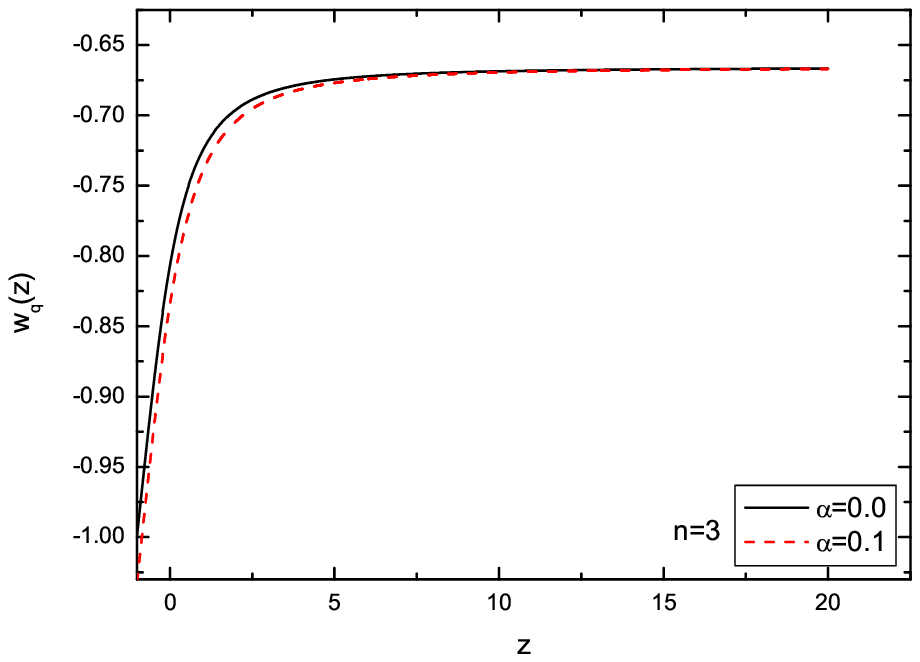}
\includegraphics[width=0.32\textwidth]{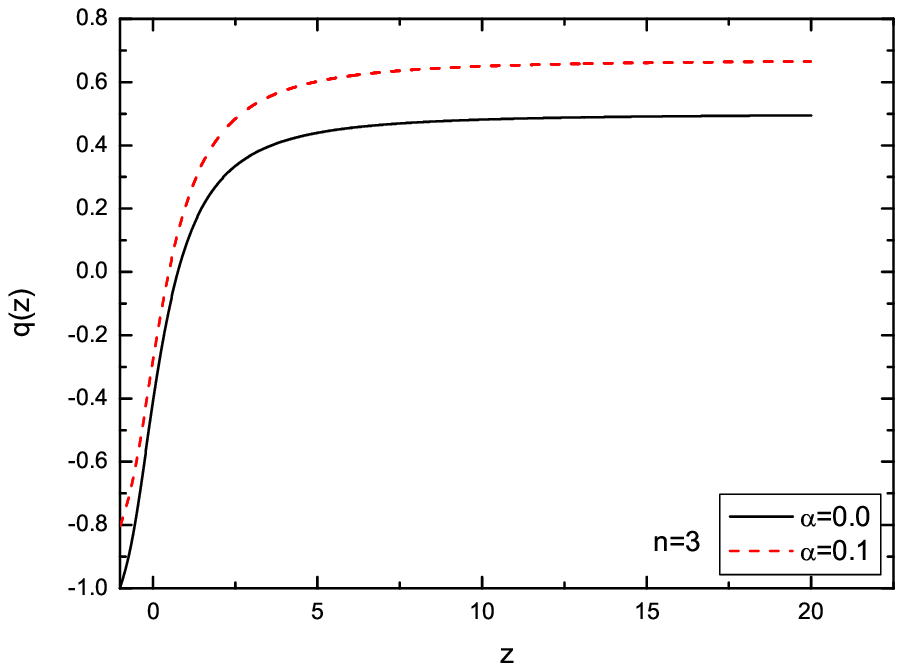}
\caption{\small The new agegraphic dark energy model in Brans-Dicke
theory: the fractional dark energy density $\Omega_{q}(z)$, the
equation of state of dark energy $w_q(z)$ and the deceleration
parameter of the universe $q(z)$. Here we use the initial condition
$\Omega_q(z_{ini})=\frac{n^2(1+\alpha)^{2}}{4(1+z_{ini})^{2}}$ at
$z_{ini}=2000$. In this figure, we first fix $\alpha=0.0001$ and
compare the cases with different $n$, and then we fix $n=3$ and
compare the usual case ($\alpha=0$) with the Brans-Dicke one
($\alpha\neq 0$).}\label{fig:2}
\end{figure}
\end{center}


To illustrate the new agegraphic dark energy model in Brans-Dicke
cosmology, we plot the fractional dark energy density
$\Omega_{q}(z)$, the equation of state of dark energy $w_q(z)$ and
the deceleration parameter of the universe $q(z)$ in
Fig.~\ref{fig:2}. The initial condition in this case is
$\Omega_q(z_{ini})=\frac{n^2(1+\alpha)^{2}}{4(1+z_{ini})^{2}}$ at
some large enough $z_{ini}$, and we choose $z_{ini}=2000$ following
Ref.~\cite{nade}. So, in the case of Brans-Dicke cosmology, the
dynamical behavior of the agegraphic dark energy is determined by
the parameters $n$ and $\alpha$. We first fix $\alpha=0.0001$ and
compare the cases with different $n$, see the upper three panels of
Fig.~\ref{fig:2}. Next, we fix $n=3$ and compare the usual case
($\alpha=0$) with the Brans-Dicke one ($\alpha=0.1$), see the lower
three panels of Fig.~\ref{fig:2}. Note that for making a distinct
comparison we take a large value of $\alpha$, namely, $\alpha=0.1$,
as the example. In the usual cosmology, the new agegraphic dark
energy behaves like a freezing quintessence \cite{agequint}, i.e.,
the equation of state $w_q>-1$ in the past and $w_q\rightarrow -1$
in the future. However, in the Brans-Dicke cosmology, the new
agegraphic dark energy behaves no more like a quintessence but like
a quintom. From Fig.~\ref{fig:2} one can see that when $\alpha\neq
0$ the equation of state of dark energy $w_q$ will cross $-1$ in the
future ($z$ approach $-1$).

\setcounter{equation}{0}
\section{New Agegraphic Dark Energy Model with
Interaction}\label{sec:5}

The interacting model of new agegraphic dark energy model in the
usual cosmology has been studied in detail in Ref.~\cite{INADE}. In
this section, we study this model in the Brans-Dicke cosmology.

Without a microscopic mechanism to characterize the interaction
between dark energy and matter, we have to use a phenomenological
term $Q$ to describe the energy exchange between dark energy and
matter, so the continuity equations can be written as
\begin{equation}
\dot{\rho_m} + 3H \rho_m=Q ,\label{eq37}
\end{equation}
\begin{equation}
\dot{\rho_q} + 3H(1+w_{q})\rho_q= -Q ,\label{eq38}
\end{equation}
which preserve the total energy conservation equation
$\dot{\rho}_{m}+\dot{\rho}_{q}+3H\left(\rho_{m}+\rho_{q}+p_{q}\right)=0$.

From Eq.~(\ref{eq27}), we get
\begin{equation}
\Omega_{q}'=\Omega_q\left(-2\frac{\dot{H}}{H^2}
-\frac{2}{a\sqrt{\tilde{n}_{0}}}\sqrt{\Omega_q}\right).\label{eq39}
\end{equation}
Using Eqs.~(\ref{eq13}), (\ref{eq25}), (\ref{eq27}) and
(\ref{eq37}), we obtain
\begin{equation}
-\frac{2\dot{H}}{H^2}=(3+\alpha)\left(1-\Omega_q\right)
+\frac{2\Omega_q\sqrt{\Omega_q}}{a\sqrt{\tilde{n}_{0}}}-\frac{2Q}{(6+6\alpha-\omega\alpha^{2})
H^3\Phi(t)}.\label{eq40}
\end{equation}
Therefore, we find that the equation of motion for $\Omega_q$ is
changed to
\begin{equation}
\Omega_{q}'={\Omega_q}
\left\{\left(1-\Omega_q\right)\left[(3+\alpha)
-\frac{2}{\sqrt{\tilde{n}_{0}}}\frac{\sqrt{\Omega_q}}{a}\right]
-\frac{2Q}{(6+6\alpha-\omega\alpha^{2})\Phi(t)H^3}\right\}.\label{eq41}
\end{equation}
Furthermore, from Eqs.~(\ref{eq38}), (\ref{eq27}) and (\ref{eq25}),
we obtain the equation of state of dark energy
\begin{equation}
w_q=-\frac{1}{3}[(3+\alpha)-\frac{2}{\sqrt{\tilde{n}_{0}}}\frac{\sqrt{\Omega_q}}{a}+\frac{Q}{H\rho_q}].\label{eq42}
\end{equation}

With the relation $\omega\alpha=6$, we have $\tilde{n}_{0}=n^2$, as
discussed previously. In addition, as a phenomenological example, we
take $Q=3\beta H(\rho_m+\rho_q)$, where $\beta$ is the coupling
constant, following the literature. So, Eqs.~(\ref{eq41}) and
(\ref{eq42}) can rewritten as
\begin{equation}
\frac{d\Omega_{q}}{dz}=-\Omega_q
\left\{\left(1-\Omega_q\right)\left[(3+\alpha)(1+z)^{-1}
-\frac{2}{n}\sqrt{\Omega_q} \right]
-3\beta(1+z)^{-1}\right\},\label{eq43}
\end{equation}
\begin{equation}
w_q=-\frac{1}{3}[(3+\alpha)-\frac{2}{n}\sqrt{\Omega_q}(1+z)+\frac{3\beta}{\Omega_{q}}].\label{eq44}
\end{equation}

\begin{center}
\begin{figure}[htbp]
\centering
\includegraphics[width=0.32\textwidth]{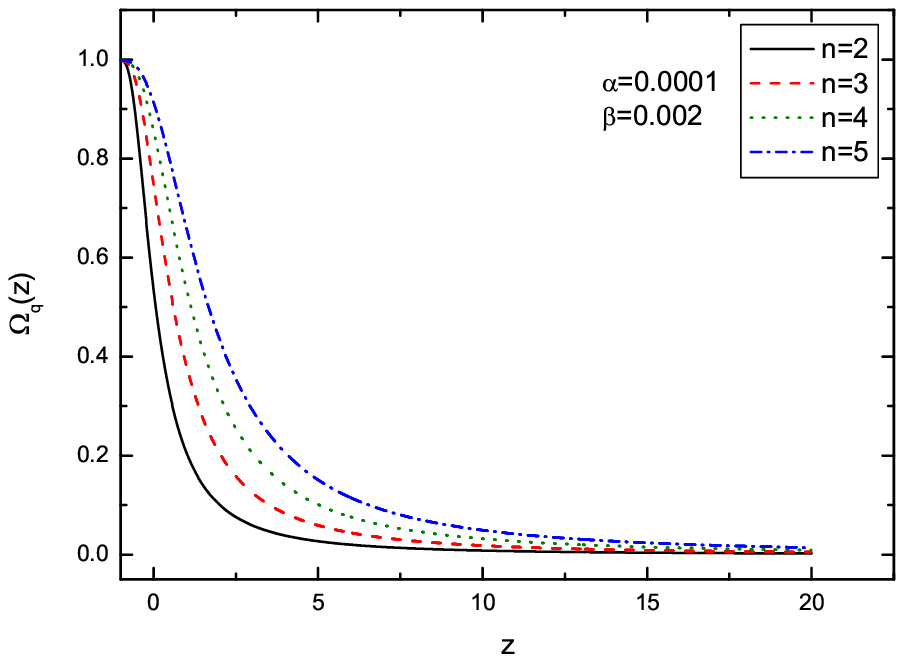}\hfill
\includegraphics[width=0.32\textwidth]{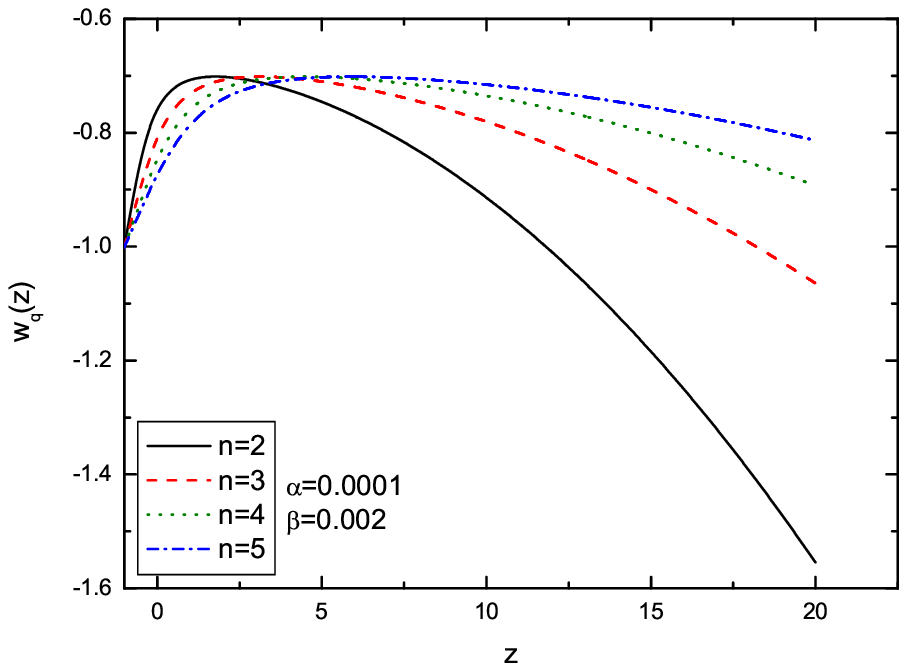}
\includegraphics[width=0.32\textwidth]{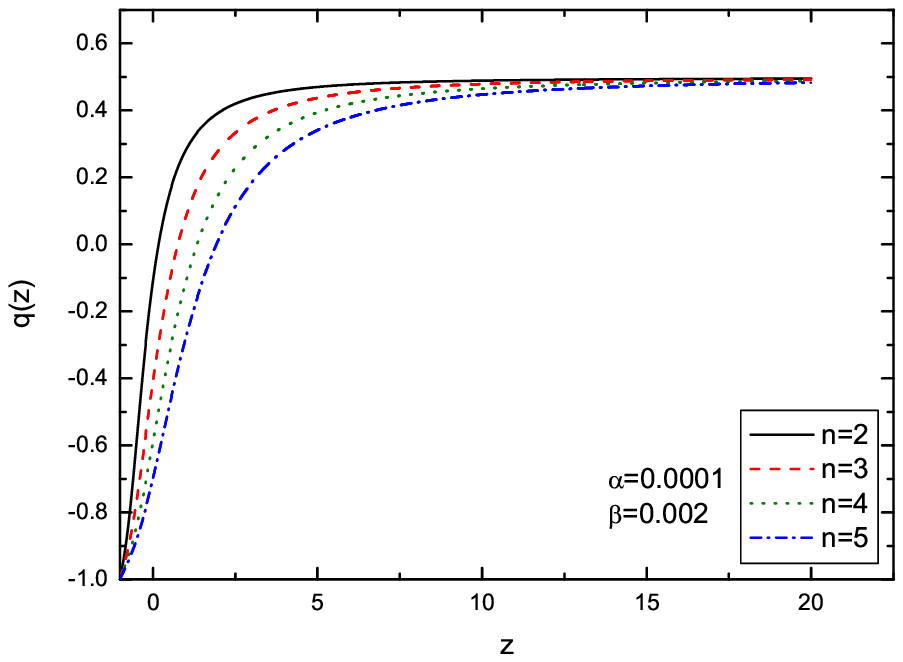}
\caption{\small The interacting model of new agegraphic dark energy
in Brans-Dicke theory. In this example, we fix $\alpha$ and $\beta$,
and vary $n$.}\label{fig:3}
\end{figure}
\end{center}
\begin{center}
\begin{figure}[htbp]
\centering
\includegraphics[width=0.32\textwidth]{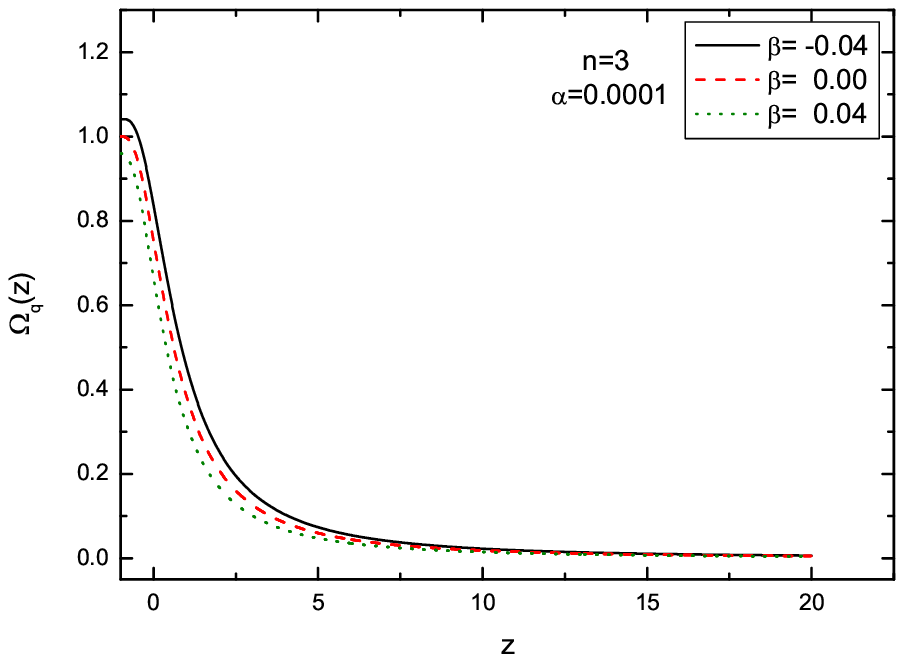}\hfill
\includegraphics[width=0.32\textwidth]{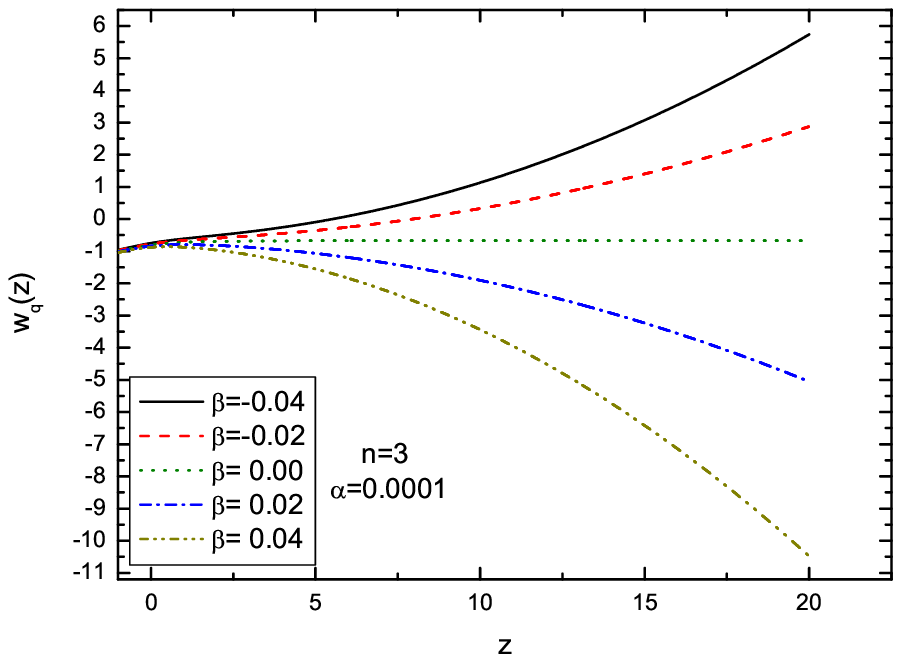}
\includegraphics[width=0.32\textwidth]{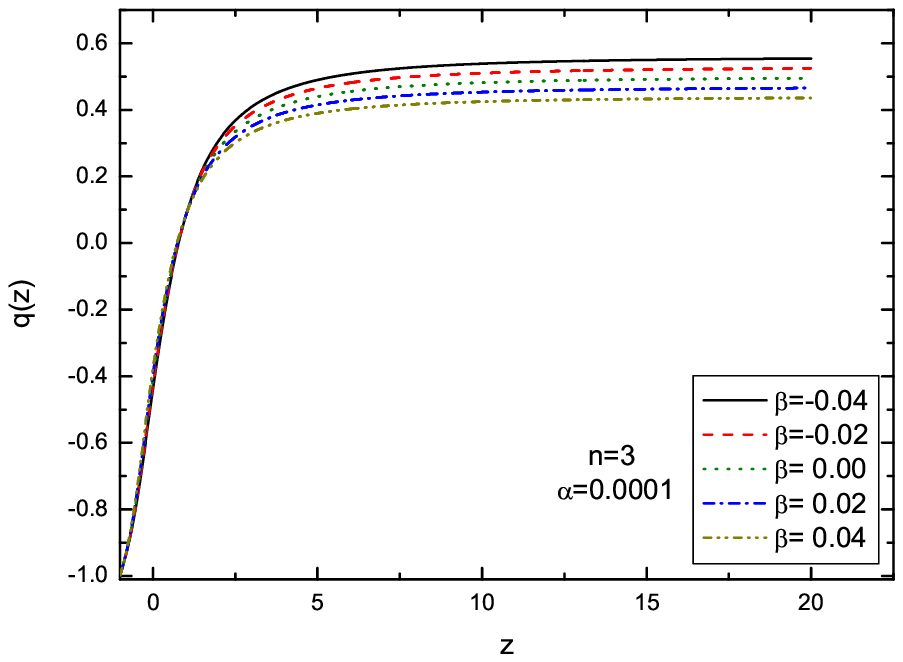}
\caption{\small The interacting model of new agegraphic dark energy
in Brans-Dicke theory. In this example, we fix $\alpha$ and $n$, and
vary $\beta$.}\label{fig:4}
 \end{figure}
 \end{center}


To illustrate the cosmological evolution of the interacting model of
new agegraphic dark energy in Brans-Dicke theory, we plot the
fractional dark energy density $\Omega_q(z)$, the equation of state
of dark energy $w_q(z)$ and the deceleration parameter $q(z)$ in
Figs.~\ref{fig:3} and \ref{fig:4}. For the interacting case, the
same initial condition
$\Omega_q(z_{ini})=\frac{n^2(1+\alpha)^{2}}{4(1+z_{ini})^{2}}$ at
$z_{ini}=2000$ can still be used, for the detailed discussion see
Ref.~\cite{INADE}. First, to see the effect of the agegraphic model
parameter $n$, as an example, we fix $\alpha=0.0001$ and
$\beta=0.002$, and vary $n$ (from 2 to 5), as shown in
Fig.~\ref{fig:3}. From this figure, it is very interesting to find
that the interaction could break the early-time degeneracy in $w_q$
for various values of $n$ (see Fig.~\ref{fig:2} for comparison).
Next, to see the effect of the interaction parameter $\beta$, as an
example, we fix $n=3$ and $\alpha=0.0001$, and vary $\beta$ (the
value is taken as $-0.04$, $-0.02$, 0, $0.02$, and $0.04$,
respectively), as shown in Fig.~\ref{fig:4}. From this figure, one
can clearly see the impact of the interaction between dark energy
and matter on the cosmological evolution of the new agegraphic dark
energy model. It can be explicitly seen from the left panel of
Fig.~\ref{fig:4} that $\beta<0$ will lead to unphysical consequences
in physics, since $\rho_m$ will become negative and $\Omega_q$ will
be greater than 1 in the future. So, $\beta=b^2$ is commonly assumed
in the literature.

\setcounter{equation}{0}
\section{Conclusion}\label{sec:6}

In this note, we study the agegraphic dark energy model in the
framework of Brans-Dicke gravitational theory. The Brans-Dicke
theory is a natural alternative and a simple generalization of the
Einstein's general relativity. It is also the simplest example of a
scalar-tensor gravitational theory. In the Brans-Dicke theory, the
gravitational constant is replaced with the inverse of a
time-dependent scalar field. We investigated how the agegraphic dark
energy evolves in the universe in such a gravitational theory. In
the Brans-Dicke cosmology, we derived the equation of state of dark
energy $w_q(z)$ and the deceleration parameter $q(z)$ in both old
and new versions of agegraphic dark energy model (in spit of the
internal inconsistencies in the old version). In the usual
cosmology, the agegraphic dark energy behaves like a quintessence:
the old agegraphic dark energy looks like a thawing quintessence and
the new agegraphic dark energy mimics a freezing quintessence.
However, it is very interesting to find that in the Brans-Dicke
theory of gravity the agegraphic dark energy (both old and new)
realizes a quintom behavior, i.e., its equation of state crosses the
phantom divide $w=-1$ during the evolution. We compared the
situation of the agegraphic dark energy model in the Brans-Dicke
theory with that in the Einstein's theory. In addition, we also
discussed the interaction model of the new agegraphic dark energy
model in the Brans-Dicke theory.

\section*{Acknowledgements}

This work was supported by the National Natural Science Foundation
of China under Grant No. 10705041.

\vskip 0.2cm

\noindent{\it Note Added:} During the submission and review process
of this manuscript, Ref. \cite{Sheykhi:2009yn} appeared on the arXiv
which discusses the similar topic to our study.


\end{document}